# Accurate determination of electron-hole asymmetry and next-nearest neighbor hopping in graphene

A. Kretinin[1], G. L. Yu[2], R. Jalil[1], Y. Cao[1], F. Withers[2], A. Mishchenko[2], M. I. Katsnelson[3], K. S. Novoselov[2], A. K. Geim[1,2], F. Guinea[4,5]

[1]Manchester Centre for Mesoscience and Nanotechnology, University of Manchester, Oxford Road, Manchester, M13 9PL, UK.
[2]School of Physics and Astronomy, University of Manchester, Oxford Road, Manchester, M13 9PL, UK.
[3]Theory of Condensed Matter, Institute for Molecules and Materials, Radboud University Nijmegen, Heyendaalseweg 135, 6525 AJ Nijmegen, The Netherlands.
[4]Instituto de Ciencia de Materiales de Madrid (ICMM-CSIC), Cantoblanco, 28049 Madrid, Spain.
[5]Donostia International Physics Center. (DIPC), P. de Manuel Lardizabal 4, 20018, San Sebastián, Basque Country, Spain.

The next-nearest neighbor hopping term $t'$ determines a magnitude and, hence, importance of several phenomena in graphene, which include self-doping due to broken bonds and the Klein tunneling that in the presence of $t'$ is no longer perfect. Theoretical estimates for $t'$ vary widely whereas a few existing measurements by using polarization resolved magneto-spectroscopy have found surprisingly large $t'$, close or even exceeding highest theoretical values. Here we report dedicated measurements of the density of states in graphene by using high-quality capacitance devices. The density of states exhibits a pronounced electron-hole asymmetry that increases linearly with energy. This behavior yields $t' \approx$ - 0.3eV ±15%, in agreement with the high end of theory estimates. We discuss the role of electron-electron interactions in determining $t'$ and overview phenomena which can be influenced by such a large value of $t'$.

The conduction and valence bands of graphene are well described by a tight binding model based on the carbon $p_z$ orbitals. The nearest neighbor term, $t \approx 3$ eV gives a reasonable description of the Dirac cones at low energies [1], with a Fermi velocity $v_F = (3ta)/(2\hbar)$ where $a \approx 1.4$ Å is the distance between nearest neighbor carbon atoms. The bands calculated by assuming only the nearest neighbor hopping are electron-hole (e-h) symmetric. If the next nearest neighbor term $t'$ is taken into account, the e-h symmetry becomes broken with the low energy electronic dispersion written as

$$\epsilon_k \approx -3t' \pm \hbar v_F k + \frac{9}{4} t'(ka)^2 + \cdots \qquad (1)$$

The position of the Dirac point is shifted by $-3t'$. The breakdown of electron-hole symmetry does not change significantly the electronic properties of the system at long wavelengths. However, the asymmetry results in a finite dispersion of edge states and changes the midgap states near a vacancy into a resonance. Also, in the presence of the asymmetry, extended defects can lead to self doping [2,3,4].

Theoretical estimates of the value of $t'$ vary [5,6]. Detailed calculations suggest that $t'$ depends on interaction effects [6]. Experimentally, the parameter has been studied by using polarization resolved magneto-spectroscopy that reveals a difference in the energy separation between Landau levels for electron and hole bands. The first such experiment [7] performed several years ago inferred $t' \approx 0.9$ eV but this high value could not be justified theoretically. However, the early devices were of low quality (graphene on silicon oxide with carrier mobility ≈10,000 cm² V⁻¹ s⁻¹). The recent experiment [8] in high magnetic fields of ~20T and using high-quality "graphene on graphite" has found $|t'| \approx 0.4$ eV, which is consistent with theories but still lies on the high side of the expected range.

**Quantum capacitance measurements and determination of $t'$**

In this work, to determine $t'$ we have employed capacitance measurements of graphene encapsulated in hexagonal boron nitride (hBN). Our devices are schematically shown in the inset of Fig. 1. They are designed similar to the graphene capacitors reported in ref. 9 but exhibit higher quality and homogeneity that are essential for the measurements' accuracy. Briefly, graphene in deposited on top of atomically flat hBN and then another hBN crystal of a small thickness (15–25 nm) is deposited on top. An evaporated gold film completes the capacitor and serves as one of its electrode. The second electrode is graphene. The differential capacitance $C$ of such devices is measured as a function of voltage bias $V$ between the two electrodes by using a capacitance bridge [9]. $C$ exhibits a typical value of 0.1-0.3 pF for our devices with an active area $S \approx$100-200 μm². Measurements in magnetic fields reveal the onset of Landau quantization at ≈0.1 T, which yields a quantum mobility of ≈100,000 cm² V⁻¹ s⁻¹. Comparison of this value with quantum mobility measured on similar encapsulated devices but with smaller $S \approx$10 μm² indicates that quantum oscillations in large graphene capacitors underestimate their mobility by a factor of a few. The disagreement can be explained by the fact that quantum oscillations are smeared not only by scattering but, also, by charge inhomogeneity that increases with increasing $S$.

The measured $C$ consists of geometrical and quantum capacitance contributions [9,10]. The former capacitance $C_G$ is independent on $V$ and can be subtracted as a single fitting parameter that has a value close to the saturation value of $C(V)$ at high $V$ [9]. The resulting quantum capacitance $C_Q = Se^2\, dn/d\mu$ can then be replotted as a function of carrier concentration $n$ rather than $V$ where $e$ is the electron charge, $\mu$ the chemical potential and, for noninteracting electrons, $\frac{dn}{d\mu} = N(\epsilon_F)$ is the density of states (DoS) at the Fermi energy. Let us emphasize that, for graphene devices with a thin gate dielectric, the usual approximation $n \propto V$ is no longer valid because $C$ becomes a function $V$ and changes by a factor of 2 for our typical devices at liquid-helium temperature. To this end, we have determined $n$ by numerically integrating the experimentally measured $C(V)$ over $V$. For further details about our experimental devices, procedures and data analysis, we refer to the earlier reports [9,10].

Figure 1a shows $C_Q(n)$ for one of our devices. The quantum capacitance varies as $4Se^2\sqrt{\pi|n|}/hv_F$, that is, approximately $\propto |\epsilon_F|$ ($h$ is Planck's constant), yielding the Fermi velocity $v_F(n)$ of about $1\times10^6$ m/s. For $|n|$ below a few $10^{11}$ cm$^{-2}$, $v_F$ shows a pronounced peak, increasing by a factor of 2, in good agreement with the previous measurements of many-body renormalization of the Dirac spectrum [9,11]. In this report, we focus on e-h asymmetry, which is practically indiscernible within our experimental accuracy in the previously studied regime of low $n$ but becomes notable at higher $n$. The red curve in Fig. 1a emphasizes this asymmetry. It is clear that the DoS for electrons $N^+$ is notably higher than that for holes $N^-$ and the difference increases with $n$. Figure 1b plots this difference $\Delta N = N^+ - N^-$, normalized to the average DoS, $<N> = (N^+ + N^-)/2$. It is instructive to present $\Delta N$ as a function of the electron and hole energy $\epsilon$ rather than $n$. In the first approximation, $\epsilon$ can be calculated as $\epsilon_F = \hbar v_F \sqrt{|n|/\pi}$, which assumes a constant $v_F$. We can also avoid this assumption by using the expression $\epsilon_F = eV - e^2 nS/C_G$ [9]. Both approaches yield behavior that is practically indistinguishable over our range of $n$ and presented in Fig. 1b. It shows that, within the experimental error, five capacitor devices studied in our work exhibit the same values of $\Delta N$, proving reliability of the results.

One can also see that the data in Fig. 1b can be fit by the linear dependence $\Delta N/N = \alpha \times \epsilon$ expected theoretically (see below). According to the tight-binding model, $\alpha$ is given by $\approx 6t'/t^2$. If we assume $t \approx 3$eV, Fig. 1b yields $|t'/t| \approx 0.1$ or $t' \approx -0.3$eV with statistical accuracy of $\pm 10\%$. However, $t$ itself has not been accurately defined ($|t| \approx 2.7$ eV has also been suggested in the literature) and this reduces the accuracy of determining $t'$ to $\pm 15\%$. This value is 3 times lower than $t'$ suggested in the early graphene work [7] and somewhat lower than $\approx 0.4$eV reported recently [8] (no accuracy was specified in the latter work).

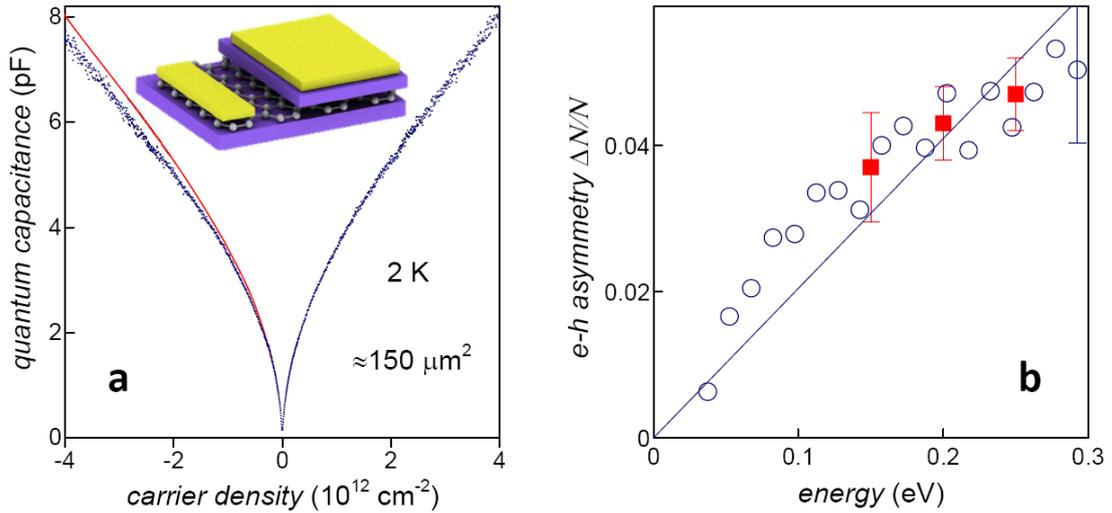

FIG 1. Asymmetry in the DoS of monolayer graphene. **a** – Quantum capacitance as a function of $n$ (positive and negative $n$ refer to electrons and holes, respectively). Dots are the experimental data. The red curve mirrors the average behavior of $C_Q$ at positive $n$ to highlights the e-h asymmetry. Inset: devices' schematic. The top electrode and the gold contact to graphene are shown in yellow; hBN is in violet. **b** – Asymmetry in the DoS. Blue circles – experimental data averaged over consecutive energy intervals of 30 meV for the device in (a). The blue line is the best linear fit. Red dots are measurements for four similar devices averaged over 100 meV and over different devices. The bars show statistical errors. Our experimental accuracy is found to be highest around $\epsilon = 0.2$eV because the asymmetry decreases with decreasing $\epsilon$ whereas the accuracy of measuring $C_Q$ also decreases with increasing $\epsilon$ because the geometrical capacitance start dominating $C$ at high $n$.

***Effect of interactions on e-h symmetry***

The long range part of the Coulomb interaction modifies the quasiparticle dispersion near the Dirac energy [11,12]. This effect can be formulated as the generation by the interaction of a self energy which has a contribution which depends logarithmically on the ratio between the high momentum cutoff of the model and the quasiparticle momentum [12]

$$\delta\epsilon_k \approx \delta_1\epsilon_k + \delta_2\epsilon_k + \cdots \approx \pm \frac{e^2}{4\epsilon_0} k \log\left(\frac{\Lambda}{k}\right) + O\left(\frac{k^2}{\Lambda}\right) + \cdots \quad (2)$$

where $k$ is the momentum, $\Lambda \approx a^{-1}$ the high $k$ cutoff in the model, $\epsilon_0$ the dielectric constant of the environment, and the signs in the first term refer to the conduction and valence bands. A $k$-independent constant, proportional to $e^2\Lambda/\epsilon_0$, has been omitted in eq. (2). In Renormalization Group language, the terms left out in eq. (2) are irrelevant, and they can be neglected in the limit $k/\Lambda \to 0$. In a condensed matter model, however, the ratio $k/\Lambda$ is typically finite, and irrelevant terms give a finite, nonsingular contribution. If we describe the interactions in terms of the bare Coulomb potential, $v_q = (2\pi e^2)/(\epsilon_0 q)$ and estimate the self energy by using the first order perturbation theory, on dimensional grounds we obtain [12]

$$\delta_2\epsilon_k = c\frac{e^2 k^2}{\epsilon_0 \Lambda} \approx \bar{t}(ka)^2 \quad (3)$$

as the only available parameters are $e^2/\epsilon_0$, $\Lambda$, and $k$. In eq.(3) $c$ is a numerical coefficient, and $\bar{t}$ is a parameter with dimension of energy.

Alternatively, we can use the Hartree-Fock approximation to estimate the self energy. By using the Coulomb potential described above and the wavefunctions derived from the Dirac approximation to the electronic bands, the self energy can be written as [13]

$$\delta\epsilon_k^{HF} = -\frac{e^2}{4\pi\epsilon_0}\int_0^\Lambda k'dk' \int_0^{2\pi} d\theta \frac{1\pm\cos\theta}{\sqrt{k^2+k'^2-2kk'\cos\theta}} \quad (4)$$

where the two signs in the integral correspond to the valence and conduction bands. This expression can be expanded in powers of $\Lambda$:

$$\delta\epsilon_k^{HF} \approx -\frac{e^2\Lambda}{2\epsilon_0} \pm \frac{e^2}{4\epsilon_0}k\log\left(\frac{\Lambda}{k}\right) - \frac{e^2 k^2}{8\pi\epsilon_0\Lambda} + \cdots \quad (5)$$

The leading $k$ dependent term gives the logarithmic correction to the self energy as per eq. (2), whereas the next term has the same form as eq. (3).

Both above approaches suggest that a self energy term which depends quadratically on momentum $k$ can arise from electron-electron interactions, namely, from high-$k$ exchange processes. It is determined by the short range features of the interaction, and the continuum approach used previously cannot be expected to be numerically accurate. Unlike the logarithmic correction to the Fermi velocity, this term is not affected by the screening properties of the environment. Using the above formulas, we can estimate an effective next nearest neighbor hopping term $t'$ as $\approx e^2/(18\epsilon_0\Lambda a^2)$. For $e^2/v_F \approx 2$ and $\epsilon_0 \approx 2$ we obtain $t' \approx -|t|/12$. This estimate is close to the value of $t'$ found experimentally.

### Influence of a large $t'$ on graphene's properties
*Defect states near the neutrality point*
In neutral graphene, there can exist localized states that appear due to defects such as edges [14-18], vacancies [3] and other confinement effects [4,19]. For $t' = 0$, the wavefunctions associated with these states are finite only on one of the graphene sublattices and, therefore, the states appear exactly at zero energy. These wavefunctions are also solutions of the local tight binding equations when $t' \neq 0$ (see Appendix). However, a finite value of $t'$ modifies the boundary conditions and shifts the localized states from the Dirac point so that they become resonances with a finite decay width.

For the case of a vacancy, its state is shifted from zero energy $\varepsilon$ by a value of the order of $\Delta\varepsilon \propto 3t'/[\log(R/a)]^{3/2}$ where $R$ is a long distance cutoff comparable to the size of a graphene device (see Appendix). In practice, charge accumulation at the sites nearest to the vacancy shifts the resonance back to zero $\varepsilon$, so that away from the vacancy site graphene remains neutral. The smallness of the shift induced by $t'$ is consistent with the experimental observation of sharp vacancy resonances [20].

For a generic graphene edge of length $L$ which also gives rise to localized states, we expect that their number is a fraction of $L/a$, and their decay lengths range between $L$ and $a$. Therefore, $\Delta\varepsilon$ due to a finite value of $t'$ should vary between $t'(a/L)$ and $t'$. The additional density of states associated with graphene edges then changes from a delta function at zero $\varepsilon$ to $D(\varepsilon)\sim(L/a)/t'$ within the energy interval $0 \leq \varepsilon \leq t'$. The resonant states at vacancy or edge sites result in uncompensated spins, which should also exhibit e-h asymmetry. To compensate the spins requires a shift of the chemical potential of at least $\sim t'$, in agreement with the recent experiment [21].

Self doping at the edges is expected to result in charge accumulation [2], which reduces the tendency towards the formation of uncompensated spins [4,18].

*Thermal self doping*
At a finite temperature $T$, the broken e-h symmetry due to $t'$ induces different amounts of electrons and holes and, therefore, induces some doping at finite temperatures. The density of states is given by

$$D(\epsilon) \approx \frac{8\epsilon}{9\pi t^2 a^2} - \frac{8t'\epsilon^2}{3\pi t^3 a^2} + \cdots \quad (6)$$

and, after some algebra, one can find the thermally induced density of states

$$n(T) \approx \frac{8Z(3)t'}{\pi t^4 a^2}(k_B T)^3 \approx 3.06 \frac{t'(k_B T)^3}{t^4 a^2} \tag{7}$$

where $Z(3)$ is Riemann's Zeta function. For $T = 300K$, we expect thermally-induced electron doping at a level of $\approx 1.2 \times 10^9 \text{cm}^{-2}$, which can probably be observed in dedicated experiments.

*Changes in Landau level structure*

In low magnetic fields, where a continuum model is valid, the correction to the wavefunctions of the Landau levels in monolayer graphene induced by $t'$ can be calculated analytically, see Appendix 2. Their energies are shifted by a term proportional to the magnetic field

$$\epsilon_n \approx \begin{cases} \frac{9a^2 t'}{4\ell_B^2} & n = 0 \\ \frac{9a^2 t'}{2\ell_B^2} n \pm \sqrt{\frac{2n v_F^2}{\ell_B^2} + \left(\frac{9a^2 t'}{4\ell_B^2}\right)^2} & n \neq 0 \end{cases} \tag{8}$$

where $\ell_B^{-1} = \sqrt{|eB|/\hbar}$ is the magnetic length (see [8,22]). The energies $\epsilon_n$ are measured with respect to the energy of the Dirac point. Note that the energy of the $n = 0$ Landau level acquires a dependence on the magnetic field. This state is localized in a single sublattice, as for $t' = 0$.

The situation is more complicated in bilayer graphene, where the low energy spectrum consists of four inequivalent Dirac points within each valley, which appear due to trigonal warping. At zero interlayer bias and very low $B$ such that $\ell_B \gg a(t^2/t_\perp t_3)$ numerous Landau levels emerge from this set of Dirac cones, where $t_\perp \approx 0.4$ eV is the interlayer hopping and $t_3$ gives the magnitude of trigonal warping [23]. There is one isotropic Dirac cone at $k = 0$, and three anisotropic cones at $k = 2t_\perp t_3/(3t^2 a)$. In the absence of next nearest neighbor hopping, the four Dirac cones lead to eight $n = 0$ Landau levels per valley. A finite value of $t'$ changes the degeneracy of the four Dirac points into a singlet and a triplet. The energy difference between them is $t'(t_\perp t_3/t^2)^2$. The Fermi velocity of the bilayer Dirac cones is of the order of $t'a$ and, therefore, the energy splitting induced by $t'$ can be resolved if $\ell_B \gg a(t^2 t'/t_\perp t_3^2)^2$. For realistic parameters, this regime is reached in $B$ of the order of a few mT.

On the other hand, the trigonal warping can be neglected at sufficiently high magnetic fields, $\ell_B \ll a(t^2/t_\perp t_3)$. In this case, there appears an eightfold degenerate Landau level near zero energy [23,24]. The degenerate set involves wavefunctions that correspond to the $n = 1$ Landau level of the electron gas, and a finite $t'$ shifts their energy, so that the eight levels are split into two quadruplets. The energy difference between the two subsets, to the lowest order in $t'$ or $B$, is

$$\Delta \epsilon_0 \approx \frac{81}{8} \frac{t' t^2 a^2}{t_\perp^2 \ell_B^4} \tag{9}$$

This effect increases as $B^2$. The splitting of the bulk Landau levels gives an avoided crossing at the edge of a sample as shown in Figure 2.

The splitting between $n = 0$ Landau levels induced by $t'$ is, however, much smaller than the splitting induced by electron-electron interactions. The latter is of the order of $e^2/(\epsilon_0 \ell_B)$ [25-29].

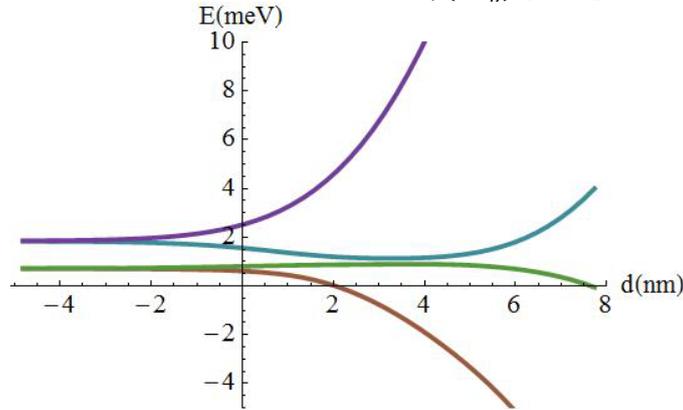

FIG 2. Dispersion of edge states for zero Landau level in bilayer graphene. $B = 25$ T; $t' = 0.4$ eV. The momentum of the state has been converted into the position of the guiding center.

*Influence of e-h asymmetry on Klein tunneling*

The next nearest neighbor term $t'$ changes the boundary conditions at interfaces, and the transmission through a ballistic p-n junction at the normal incidence becomes less than unity [30]. In a continuum model, the matching conditions at the p-n junction require the presence of evanescent waves with an inverse decay length $\lambda \approx t/(t'a) \approx 10a^{-1}$. This estimate implies that calculations of Klein tunneling in the presence of $t'$ require numerical calculations using a discrete model.

The full calculation shown in Fig. 3 necessitates the inclusion of two evanescent waves in the classical forbidden region, a situation reminiscent of Klein tunneling through p-n junctions in bilayer graphene [30,31].

One can see that the effect of a finite $t'$ is relatively small for realistic values of $t'$ and even in the case of high barriers. The new evanescent waves have decay lengths much shorter than the lattice spacing, and cannot induce major changes, either in the abrupt [30] or adiabatic [32] barrier limit.

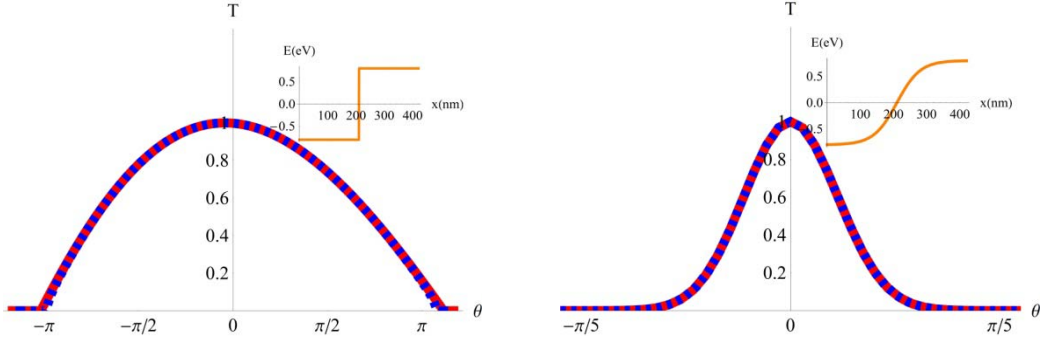

FIG 3. Transmission through a p-n junction in monolayer graphene as a function of incident angle. The chemical potentials on the left and right sides are -0.8 eV and 0.8 eV (see the insets). The red and blue (dashed) curves correspond to $t' = 0.01$ eV (effectively, zero $t'$) and $t' = 0.3$ eV. Left: abrupt barrier with a width of 2 Å. Right: smooth, 100-nm wide barrier. The barrier shapes used in the modeling are shown in the insets.

*Optical absorption in the presence of e-h asymmetry*
Light absorption is determined by optical conductivity of graphene, which depends on the velocity-velocity correlations at finite frequencies and small wavevectors $k$ [33]. A finite $t'$ modifies the velocity operator that in the continuum limit becomes

$$\hat{v}_F(\vec{k}) \equiv v_F \vec{\sigma} + \frac{9\,t'a^2}{2}\vec{k}\begin{pmatrix}1 & 0\\ 0 & 1\end{pmatrix} \quad (10)$$

where the velocity is expressed as an operator in the sublattice basis. The conductivity for $\vec{k} \to 0$ becomes

$$\sigma = \frac{e^2}{h}\frac{2}{\pi}\frac{v_F^2}{\omega}\int k dk\, \delta(\omega - \epsilon_k^c + \epsilon_k^v) = \frac{e^2}{h}\frac{2}{\pi}\frac{v_F^2}{\omega}\int k dk\, \delta(\omega - 2v_F k) = \frac{e^2}{h} \quad (11)$$

where $\epsilon_k^c$ and $\epsilon_k^v$ refer to the conduction and valence band. The contribution from $t'$ to each quasiparticle energy vanishes when the difference is computed in eq. (11) and, therefore, $t'$ does *not* induce any change in the optical conductivity at any frequency for the asymmetric conical spectrum. This result explains the excellent agreement between the experimentally found $\sigma \approx \frac{e^2}{h}$ at visible frequencies and the simple theory that did not take into account the e-h asymmetry [34]. Similarly, $t'$ does not change the energy difference between the saddle points of the $p_z$ bands located at the $M$ points of the Brillouin zone, although optical transitions between these states can be influenced by interactions within the electron-hole pair created by the photon [35,36].

*Plasmons*
Changes in the Fermi velocity due to $t'$ lead to changes in the plasmon dispersion described by

$$h\omega_p(q) = \sqrt{\frac{2e^2 h v_F(k_F) k_F q}{\epsilon_0}} \quad (12)$$

where $v_F(k_F) \approx 3ta/(2h) \pm 9t'a^2 k_F/(2h)$ and the different signs correspond to electrons and holes. Accordingly, the dependence of $\omega_p$ on $n$ changes and the plasmon frequencies become somewhat different for electrons and holes at the same density $n$.

*Electronic susceptibility, magnetic impurities and RKKY interactions*
For brevity, we consider here charge and spin susceptibilities at the neutrality point. For $t' \neq 0$ the charge susceptibility is

$$\chi_\rho(q) = \frac{q}{\pi^2 v_F}\int_0^{2\pi} d\theta \int_0^{\Lambda/q}\left(1 - \frac{k^2 - 1/4}{\sqrt{(k^2+1/4)^2 - k^2 \cos^2\theta}}\right) \times \quad (13)$$

$$\times \frac{k dk}{\sqrt{k^2 + 1/4 + k\cos\theta} + \sqrt{k^2 + 1/4 - k\cos\theta} + (9t'a^2 kq \cos\theta)/(4v_F)}.$$

For $t' = 0$ this expression gives $\chi_\rho(q) = q/(4v_F)$. From eq. (13) it is clear that the deviations induced by $t'$ become important for $q \geq v_F/t'a^2$, which is larger than $\Lambda$. A numerical integration of eq. (13), setting $\Lambda = \infty$ is shown in Fig. 4.

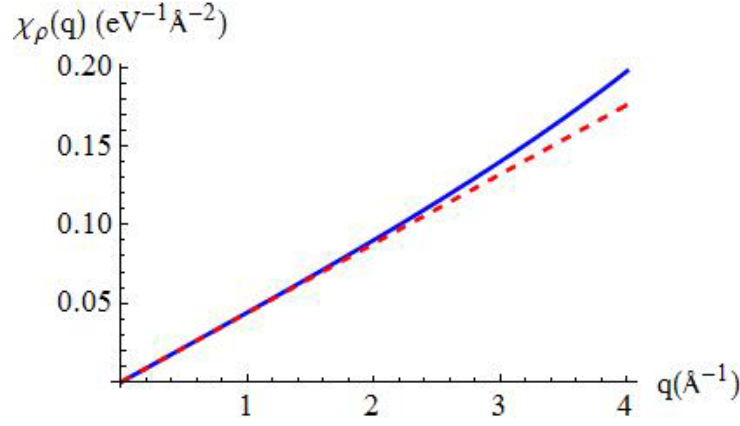

Fig. 4. Static charge susceptibility of graphene for $t' = 0$ (red, dashed) and for $t' = 0.4$ eV (blue, full).

The RKKY interactions between fixed magnetic moments at different sites in the graphene lattice can be obtained from the Fourier transform of the charge susceptibility [37,38]. The factor in parenthesis in eq. (1) has to be replaced by 1 for the interaction between moments at sites in the same sublattice, and by $-(k^2 - 1/4)/\sqrt{(k^2 + 1/4)^2 - k^2 \cos\theta^2}$ for sites in different sublattices. As for $t' = 0$ the RKKY interaction is ferromagnetic for spins in the same sublattice and antiferromagnetic for sites in different sublattices. In both cases the interaction decays as $R^{-3}$, where $R$ is the distance between spins.

Finally, the formation of magnetic moments and the Kondo effect is sensitive to the hybridization between magnetic impurities and the graphene bands. This hybridization depends essentially on high energy features of the band structure, like the position of the van Hove singularities [38], which are affected by the value of $t'$. As a result, the concentration dependence of the Kondo temperature should exhibit a strong e-h asymmetry even for relatively small $t'$ as follows from calculations for Co on graphene [39].

*Multilayer graphene*
Bilayer and Bernal-stacked graphene exhibit parabolic bands at low energies with a dispersion relation $\epsilon_k \approx (hv_F k)^2/t_\perp$. The parameter $t'$ induces another quadratic term in eq. (1), which however has a smaller magnitude. For the case of rhombohedral or ABC stacked graphene, the simplest approximation gives a low energy band with the dispersion relation $\epsilon_k \approx (hv_F k)^N/t_\perp^{N-1}$, where $N$ is the number of layers [40]. The inclusion of the interlayer hopping term $\gamma_4 \approx 0.04$ eV modifies this dispersion [41], which becomes, to the lowest order, $\epsilon_k \approx 3hv_F\gamma_4 ak^2/t_\perp$. For the found large $t' \approx 0.3$ eV, the contribution arising from eq. (1) is comparable to the leading effect of interlayer coupling.

*Effects of strain on next nearest neighbor hopping*
Lattice deformations modify the tight binding parameters, and the hybridization between orbitals, leading to changes in the electronic structure [42]. The $t'$ term leads to the appearance of a scalar potential if the graphene layer is curved [43]. In-plane strains change the interatomic distances and can also modify the value of $t'$. As a result, a different scalar potential is induced

$$V(\vec{r}) = -3\ t'\beta'[u_{xx}(\vec{r}) + u_{yy}(\vec{r})] \quad (13)$$

where $\beta' = \frac{a'}{t'}\frac{\partial t'}{\partial a'}$, $a'$ is the distance between next nearest neighbor atoms and $u_{ij}$ is the strain tensor. There is a significant uncertainty regarding the scalar potential induced by strains [44-46]. For $\beta' \approx 3$ we find a prefactor in eq. (13) of approximately 4 eV, which is consistent with the calculations in ref. 46.

*Other effects of finite $t'$*
When studying localization phenomena in graphene with different kinds of disorder, the chiral symmetry intimately related with the e-hole symmetry leads to important consequences determining the choice of the universality class [47]. Even relatively weak violation of these symmetries may in principle lead to important consequences, especially in the regime of low conductivity, that is, in the vicinity of neutrality point [38]. Furthermore, the e-h symmetry or its absence can be crucial for nonlinear optics phenomena in graphene such as second harmonic generation in the presence of valley polarization [48].

*Conclusion*
Capacitance measurements reported here yield the next nearest neighbor hopping in graphene $t' \approx -0.3$ eV, which is close to the upper bound of theoretical estimates [49]. The sign and value of $t'$ are consistent with estimates for the contribution from electron-electron interactions to the quasiparticle self energy, expanding beyond the lowest order approximation.

The $t'$ term breaks the symmetry between electrons and holes as seen directly in the measured density of states in graphene. However, this asymmetry leads to relatively weak effects in the optical properties of graphene, its electronic transport and plasmonics. The effect of $t'$ on other low energy phenomena, such as spin transport, is also expected to be negligible. Nonetheless, the additional hopping can induce nontrivial self doping effects. In particular, resonances near lattice defects become shifted from zero energy, giving rise to bands of quasi-localized states with an energy width of $t'$. The hopping term is also expected to modify strongly the band structure of graphene multilayers with rhombohedral stacking. The dependence of $t'$ on lattice deformations is consistent with estimates for the changes in chemical potential induced by strain.

Finally, let us note that e-h symmetry breaking terms, similar to $t'$ considered here, can be expected near secondary Dirac points induced by superlattices [50-52] and in artificially engineered Dirac systems[53,54].

### Acknowledgements
We acknowledge financial support from the European Research Council, the Royal Society, the Spanish Ministry of Economy (MINECO) through Grant no. FIS2011-23713, the European Research Council Advanced Grant (contract 290846) and from European Commission under the Graphene Flagship. contract CNECT-ICT-604391.

### Appendix 1

*Structure of the wavefunction of localized states at the Dirac energy.*
In the absence of nearest neighbor hopping, $t' = 0$, the localized states which might exist in graphene at the Dirac energy have amplitude in one sublattice only.

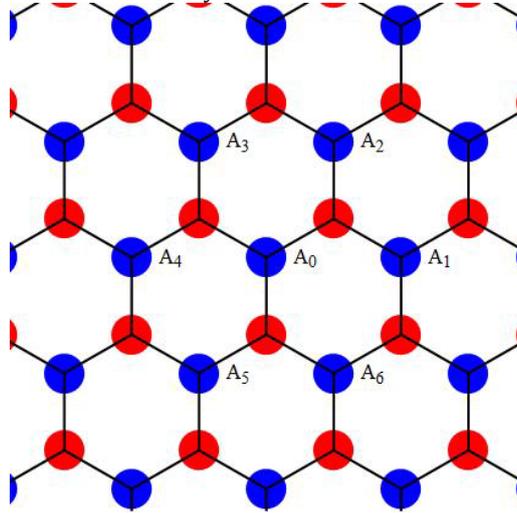

Fig. A1. Notation used for the amplitudes of states at the Dirac energy.

For $t' \neq 0$ localized states can still be defined at the Dirac energy, $\varepsilon_D = -3\,t'$, which are also defined in one sublattice only. Following the notation in Fig. A1, the amplitudes at the sites in the blue sublattice satisfy, for $t' = 0$:

$$t(A_0 + A_1 + A_2) = 0$$
$$t(A_0 + A_3 + A_4) = 0 \qquad (A1)$$
$$t(A_0 + A_5 + A_6) = 0$$

so that

$$t'(A_1 + A_2 + A_3 + A_4 + A_5 + A_6) = -3t'A_0 \qquad (A2)$$

which is the new tight binding equation induced when $t' \neq 0$ and the energy is that of the Dirac point, $\varepsilon_D = -3\,t'$.

*Resonance near a vacancy*
Equations (A1) are not satisfied in the neighborhood of a vacancy. Instead, we obtain (see notations in Fig. A2):

$$A_1 = A_2 = A_3 = \frac{A}{3}$$
$$A_1 + A_2 + A_3 = A \neq 0 \qquad (A3)$$

and, as a result,

$$t'(B_1 + B_2 + B_3 + A_2 + A_3 + B_9) = 0 \qquad (A4)$$

this equation is the tight binding equation associated with a shift of the energy at site $A_1$ by $\Delta\varepsilon_{A_1} = 9t'/A$. Similar equations can be written for sites $A_2$ and $A_3$. The constant $A$ fixes the normalization of the wavefunction, $|A|^2 \propto 1/\log(R/a)$, where $R$ is a long distance cutoff comparable to the dimensions of the graphene flake, and $a$ is a length of order of the interatomic distance.

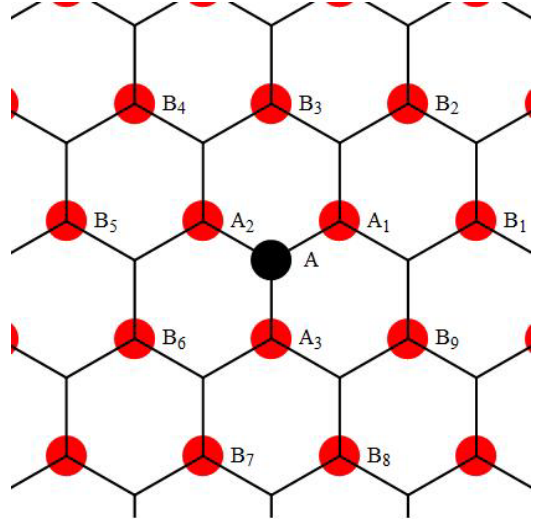

Fig. A2. Notation used for the amplitudes at the sites around a vacancy.

The previous analysis shows that a quasilocalized state near a vacancy at the Dirac energy can be defined, for $t' \neq 0$ if the three sites around the vacancy are shifted by an energy $\Delta\varepsilon \propto 9t'/\sqrt{\log(R/a)}$. The case when $\Delta\varepsilon = 0$ can be seen as a weak perturbation. Then, for $\Delta\varepsilon = 0$ we expect a resonance shifted from the Dirac energy by $\Delta\varepsilon \propto 3t'|A|^2/\sqrt{\log(R/a)} \propto 3t'/[\log(R/a)]^{3/2}$.

A similar analysis can be extended to localized states near a zigzag edge, and one obtains the results presented in ref. 14.

## *Appendix 2.*

*Influence of $t'$ on the Landau levels in monolayer graphene.*
In the continuum limit, we can use a basis of Landau levels defined in a given sublattice:

$$\Psi_{An} \equiv \begin{pmatrix} |n\rangle \\ 0 \end{pmatrix}$$
$$\Psi_{Bn} \equiv \begin{pmatrix} 0 \\ |n\rangle \end{pmatrix} \tag{A5}$$

The Hamiltonian for a given corner of the Brillouin Zone is:

$$H = \begin{pmatrix} \frac{9a^2 t'}{2\ell_B^2}\left(b^\dagger b + \frac{1}{2}\right) & \frac{\sqrt{2}v_F}{\ell_B} b \\ \frac{\sqrt{2}v_F}{\ell_B} b^\dagger & \frac{9a^2 t'}{2\ell_B^2}\left(b^\dagger b + \frac{1}{2}\right) \end{pmatrix} \tag{A6}$$

where $b^\dagger |n\rangle = \sqrt{n+1}|n+1\rangle$. The wavefunction for a given Landau level of energy $\epsilon_n$ and $n \neq 0$ can be written as

$$\Psi_n \equiv \alpha_n \begin{pmatrix} |n-1\rangle \\ 0 \end{pmatrix} + \beta_n \begin{pmatrix} 0 \\ |n\rangle \end{pmatrix} \tag{A7}$$

with

$$\begin{pmatrix} \frac{9a^2 t'}{2\ell_B^2}\left(n - \frac{1}{2}\right) & \frac{\sqrt{2}v_F}{\ell_B}\sqrt{n} \\ \frac{\sqrt{2}v_F}{\ell_B}\sqrt{n} & \frac{9a^2 t'}{2\ell_B^2}\left(n + \frac{1}{2}\right) \end{pmatrix} \begin{pmatrix} \alpha_n \\ \beta_n \end{pmatrix} = \epsilon_n \begin{pmatrix} \alpha_n \\ \beta_n \end{pmatrix} \tag{A8}$$

Expressions for $\epsilon_n$ are given in eq.(8). Note that, for $t' = 0$ and $n \neq 0$, we have $\alpha_n = \pm\beta_n = \pm 1/\sqrt{2}$. The wavefunction for the Landau level with $n = 0$ is

$$\Psi_0 \equiv \begin{pmatrix} 0 \\ |0\rangle \end{pmatrix} \tag{A9}$$

This expression is unchanged by $t'$. The velocity operator, needed for the calculation of the strength of optical transitions, becomes

$$\hat{v}_x \equiv v_F \begin{pmatrix} 0 & 1 \\ 1 & 0 \end{pmatrix} + \frac{9\sqrt{2}t'a^2}{2\hbar\ell_B} \begin{pmatrix} \frac{b^\dagger+b}{2} & 0 \\ 0 & \frac{b^\dagger+b}{2} \end{pmatrix} \tag{A10}$$

$$\hat{v}_y \equiv v_F \begin{pmatrix} 0 & -i \\ i & 0 \end{pmatrix} + \frac{9\sqrt{2}t'a^2}{2\hbar\ell_B} \begin{pmatrix} \frac{b^\dagger-b}{2i} & 0 \\ 0 & \frac{b^\dagger-b}{2i} \end{pmatrix} \tag{A11}$$

The strength of optical transitions between states $n$ and $n' = \pm(n+1)$ is proportional to factors of order $\left| \pm v_F \alpha_n \beta_{n'}^* + \frac{9\sqrt{2}t'a^2}{2\hbar\ell_B} \left( \alpha_n \alpha_{n'} \sqrt{n'} \pm \beta_n \beta_{n'} \sqrt{n'+1} \right) \right|^2$.